\documentclass[aps,pra,reprint,groupedaddress,showpacs]{revtex4-1}
\usepackage{graphicx}
\usepackage{ulem}

\begin{document}


\title{Localization, quantum resonances and  ratchet acceleration in a periodically-kicked $\mathcal{PT}$-symmetric quantum rotator}


\author{Stefano Longhi}
\email[corresponding author's email: ]{longhi@fisi.polimi.it}
\affiliation{Dipartimento di Fisica and IFN-CNR, Politecnico di Milano, Piazza L. da Vinci 32, I-20133 Milan, Italy}


\begin{abstract}
We consider wave transport phenomena in a $\mathcal{PT}$-symmetric extension of the periodically-kicked quantum rotator model and reveal that dynamical localization assists the unbroken $\mathcal{PT}$ phase. 
In the delocalized (quantum resonance) regime, $\mathcal{PT}$ symmetry is always in the broken phase and ratchet acceleration arises as a signature of unidirectional non-Hermitian transport. An optical implementation of the periodically-kicked $\mathcal{PT}$-symmetric Hamiltonian, based on transverse beam propagation in a passive optical resonator with combined phase and loss gratings, is suggested to visualize acceleration modes in fractional Talbot cavities.
\end{abstract}
\maketitle

\section{Introduction}
Since more than three decades, periodically-kicked quantum systems and related models are attracting a continuous interest as testbeds to study quantum and wave chaos, quantum resonances,  and unusual wave transport phenomena in different areas of physics \cite{r1,r2,r3,r4,r5,r6,r7,r8,r9,r10,r11,r12,r13,r14,r14bis,r15,r16,r17,r17bis,r18,r19,r20,r21}. A paradigmatic example is provided by the celebrated kicked rotor (KR) model,  which exhibits the phenomenon
of dynamical localization (DL), i.e. the quantum suppression of classical diffusion taking place in momentum space as a result of
wave interference similar to Anderson localization of the electronic wave function in disordered solids \cite{r3,r4,r5}. DL has been observed in a wide variety of physical systems, such as cold atoms \cite{r9,r10}, optical structures \cite{r13,r14,r14bis,r15,r17,r20}, and laser-driven molecules \cite{r21}. An additional quantum behavior in the KR model is the existence of accelerator modes, which arise in the presence of an external constant force like the gravity field in periodically-kicked cold atoms \cite{r11,r12,r21bis,r21tris,r21quatris}.  In such systems, a fraction of the atoms falling under the action of gravity are steadily accelerated away from the bulk of the atomic cloud, with an acceleration that can be externally controlled and may even be opposite in direction to gravity. Quantum accelerator modes do not have any counterpart in the classical dynamics and are a purely quantal (interference) phenomenon.\par
Most of previous works on quantum chaos and wave transport in periodically-kicked systems have been focused to Hermitian models. Recently, 
 new classes of chaotic systems, showing parity-time ($\mathcal{PT}$) symmetry, have been investigated \cite{r22,r23}. $\mathcal{PT}$-symmetry, originally introduced in quantum physics as a complex extension of quantum mechanics \cite{r24,r24bis}, has provided in recent years a fruitful concept in different ares of physics, ranging from optics  \cite{r25,r26,r27,r28,r28bis,r29,r30,r31,r32,r32tris} to atomic vapors and ultracold atoms \cite{r32bis,r33,r34,r34bis,r34tris,r35,r36}, acoustics \cite{r37,r38,r39}, optomechanics \cite{r23,r40}, and nonlinear physics \cite{r40bis,r41,r42,r42b}. A $\mathcal{PT}$-symmetric extension of the celebrated KR model, with an imaginary (gain and loss) gradient added to the periodic kicked potential,  has been introduced in Ref.\cite{r22}, where it was shown that chaos (i.e. absence of DL) assists the exact $\mathcal{PT}$ phase.\\
  In this article we consider a different $\mathcal{PT}$ extension of the KR model, in which the particle is periodically-kicked by  a {\it complex crystals} \cite{r42bis}. 
  Complex crystals are spatially-periodic potentials with a nonvanishing imaginary part, which have been introduced and experimentally
realized for matter \cite{r43,r44,r45,r46}  and optical \cite{r29,r30,r32} waves. Transport properties in such crystals have attracted great attention in recent years \cite{r26,r27,r29,r30,r44,r45,r46,r47,r48,r49,r50,r51,r52,r53,r54,r55,r55bis,r56}.
 As compared to ordinary crystals, wave transport in complex
crystals may exhibit some unusual properties, such as violation
of the Friedel's law of Bragg scattering \cite{r44,r45}, unidirectional scattering \cite{r27,r30,r48,r51,r52}, giant Goos-H\"{a}nchen shift \cite{r50}, Talbot revivals \cite{r53}, and unidirectional robust transport \cite{r57,r58}.\\
   The $\mathcal{PT}$ extension of the KR model considered in the present article concerns with a quantum particle which is periodically kicked by a complex crystal, notably by the $\mathcal{PT}$-symmetric sinusoidal optical potential \cite{r26,r48,r49}. The main results of the analysis are that dynamical localization assists the unbroken $\mathcal{PT}$ phase, whereas in the delocalized regime (quantum resonances) we disclose a novel kind of accelerated modes, {\it non-Hermitian accelerator modes}, which are a signature of non-Hermitian unidirectional transport. We also suggest a physical implementation in optics of the periodically-kicked $\mathcal{PT}$-symmetric model, which is based on transverse beam dynamics in a passive optical resonator with combined phase and loss gratings. In particular, we show that non-Hermitian ratchet acceleration can be observed in the far-field pattern of transient field decay when the cavity length is tuned to a fractional Talbot distance. 
 
\section{Periodically-kicked $\mathcal{PT}$ quantum rotator: Model}  
We consider an extension of the quantum KR model \cite{r3,r7}, the $\mathcal{PT}$ KR, which is described by the time-periodic Hamiltonian
\begin{equation}
\hat{H}(t)=-\frac{\hbar^2}{2I}\frac{\partial^2}{\partial x^2}+ f(t) V(x)
\end{equation}
where 
\begin{equation}
f(t)=\sum_n \delta (t-nT)
\end{equation}
 describes the periodic sequence of kicks at time intervals $T$, and $V(x+a)=V(x)$ is a complex crystal, i.e. a complex periodic potential with spatial period $a$. An important example of complex crystal is provided by the $\mathcal{PT}$-symmetric sinusoidal potential \cite{r26,r48,r49}
\begin{equation}
V(x)=V_0 \left[ \cos(2 \pi x/a)+i \lambda \sin ( 2 \pi x /a) \right]
\end{equation}
where $\lambda \geq 0$ is the non-Hermitian parameter that measures the strength of the imaginary part of the potential. While our analysis will be mostly focused to the $\mathcal{PT}$ sinusoidal potential, the main results are expected to be rather general to other complex crystals. We note that the ordinary (Hermitian) quantum KR model is obtained by assuming $\lambda=0$ and $a= 2 \pi$ in Eq.(3). On the other hand, for $f(t)=1$ the Hamiltonian $\hat{H}$ is time-independent and its energy  spectrum and corresponding Bloch eigenfunctions have been investigated in several previous papers \cite{r26,r42bis,r48,r49} : the energy spectrum is entirely real for $\lambda <1$ (unbroken $\mathcal{PT}$ phase), whereas band merging and complex energy spectrum arise for $\lambda>1$ (broken $\mathcal{PT}$ phase).  Here we will limit to consider the $\lambda<1$ region, corresponding to the unbroken $\mathcal{PT}$ phase in the non-kicked (i.e. $f(t)=1$) limit. The dynamics of $\mathcal{PT}$ KR Hamiltonian can be at best studied in momentum space. After expanding the vector state $| \psi(t) \rangle$ of the particle as $|\psi(t) \rangle= \sum_{l=-\infty}^{\infty} \psi_l(t) \exp( 2 \pi i l x/a)$, the evolution equations for the amplitude probabilities $\psi_l$ read explicitly
\begin{equation}
i \frac{d \psi_l}{dt}= \frac{2 \pi \beta l^2}{T}  \psi_l+ \frac{f(t)}{\hbar} \sum_{n=-\infty}^{\infty} V_{l-n} \psi_n 
\end{equation} 
where $V_n$ are the Fourier coefficients of the potential $V(x)$, i.e. $V(x)=\sum_n V_n \exp( 2 \pi i n x / a)$, and where we have set
\begin{equation}
\beta \equiv \frac{\pi \hbar T}{Ia^2}.
\end{equation}
\section{Quasi energy spectrum and $\mathcal{PT}$ symmetry breaking}
For the sinusoidal $\mathcal{PT}$ potential and $f(t)=1$, the energy spectrum of $\hat{H}$ is real for $\lambda<1$, whereas $\mathcal{PT}$ symmetry breaking arises at $\lambda=1$ owing to the appearance of exceptional points \cite{r48}. When the particle is periodically-kicked by the sinusoidal potential, the energy spectrum is replaced by the quasi energy (Floquet) spectrum $\epsilon$, which is defined from the eigenvalue problem
\begin{equation}
\hat{U} | \phi \rangle = \exp(-i \epsilon T) | \phi \rangle
\end{equation} 
where 
\begin{equation}
\hat{U}=\exp \left[-\frac{ i V(x)}{ \hbar} \right] \exp \left( i \frac{\beta a^2}{ 2 \pi}  \frac{\partial^2}{\partial x^2}  \right)
\end{equation}
is the time-ordered propagator $\hat{U}=\exp [-(i/ \hbar) \int_0^T dt \hat{H}(t)]$ over one period $T$. The real part of the quasi energy $\epsilon$ is defined apart from integer multiplies than $\omega= 2 \pi / T$. In momentum space, $|\phi \rangle= \sum_l \phi_l \exp( 2 \pi i l x/a)$, the quasi energy $\epsilon$ can be computed from the matrix eigenvalue problem
\begin{equation}
\exp(-i \epsilon T) \phi_l= \sum_{n=-\infty}^{\infty} \mathcal{U}_{l,n} \phi_n
\end{equation} 
where $\mathcal{U}_{l,n}$ in the matrix representation of $\hat{U}$ in momentum space 
\begin{equation}
\mathcal{U}_{l,n}=W_{l-n} \exp(-2 \pi i \beta n^2)
\end{equation}
and $W_n$ are the Fourier coefficients of $\exp[-i V(x) / \hbar]$, i.e. $\exp[-i V(x) / \hbar]=\sum_n W_n \exp ( 2 \pi i n x/a)$. Owing to the periodicity of the phase term in Eq.(9), one can limit to consider $\beta$ varying in the range $(0, 1)$. The eigenvalues $\exp(-i \epsilon T)$ of the matrix $\mathcal{U}_{l,n}$ can be numerically computed after matrix truncation, by assuming the indices $l$ and $n$ to vary in the range $l,n=-N_s,...0,1,2,..., N_s$, with a given (and possibly large) value of $N_s$. Localization (or delocalization) of the Floquet eigenstates $| \phi \rangle$ can be numerically checked by computation of the participation ratio $R=(\sum_l |\phi_l |^2)^2/\sum_l |\phi_l|^4$. For localized states $R \sim 1$, whereas for fully delocalized states $R \sim 2 N_s$. 
\par
 Let us first briefly recall some general properties of the quantum KR model in the Hermitian $\lambda=0$ limit, for which the quasi energy spectrum $\epsilon$ is real. In this case, it is known that a qualitatively different behavior is found depending on whether $\beta$ is a rational or an irrational number. The rational values of $\beta$,  $\beta=N/M$ with $N \leq M$ and $N,M$ coprime integers, correspond to so-called quantum resonances, at which the quasi energy spectrum is absolutely continuous and composed by $M$ quasi energy bands with delocalized Floquet eigenstates $| \phi \rangle$. In such a case an initially-localized state in momentum space, i.e. $\psi_l(0)=\delta_{l,0}$ in Eq.(4), will fully delocalize, i.e. the standard deviation of the momentum distribution 
\begin{equation}
\langle \Delta l (t) \rangle=\sqrt{\frac{\sum_l l^2 | \psi_l(t)|^2}{\sum_l  |\psi_l(t)|^2}}
\end{equation}
 secularly grows with time like $\sim t $. An exception occurs at $\beta=1/2$, corresponding to so-called 'anti-resonance', where collapse (flattening) of the two quasi energy bands arises and the particle state becomes trapped. On the other hand, for generic irrational values $\beta$ the phenomenon of DL is found: the Floquet eigenstates $| \phi \rangle$ are exponentially localized in momentum space and a saturation of the growth of $\langle l^2 (t) \rangle$ is observed after some time, contrary to the classical diffusive behavior. The localization length $\xi_L$ of the Floquet eigenstates for irrational $\beta$, $\phi_l \sim \exp(-|l-l_0|/ \xi_L)$, is roughly estimated to be given by \cite{r3,r4,r5,r7,r19}
 \begin{equation}
 \xi_L \sim \frac{V_0^2}{4 \hbar^2}
 \end{equation}
 which is fairly accurate as $V_o T \gg 1$.\\
   \begin{figure}[b]
\includegraphics[width=8.2cm]{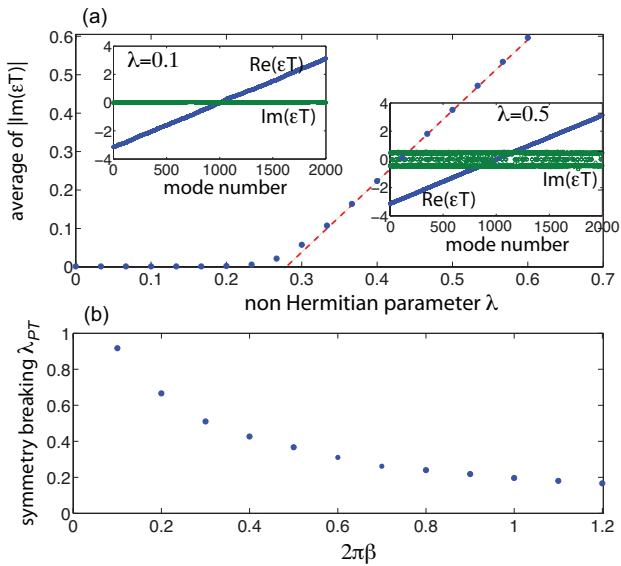}
\caption{Quasi energy spectrum $\epsilon$ and $\mathcal{PT}$ symmetry breaking threshold in the irrational $\beta$ case. (a) Numerically-computed behavior of $|{\rm Im} (\epsilon T)|$, averaged over the $2N_0+1=2001$ Floquet modes, versus $\lambda$ for parameter values $V_0 / \hbar=3$ and $ 2 \pi \beta=0.7$. $\mathcal{PT}$ symmetry breaking is clearly observed at $\lambda_{\mathcal{PT}} \simeq 0.27$. The insets show the detailed behavior of the real and imaginary parts of the quasi energies $\epsilon$ for $\lambda=0.1$ (unbroken $\mathcal{PT}$ phase) and $\lambda=0.5$ (broken $\mathcal{PT}$ phase). (b) Numerically-computed behavior of the $\mathcal{PT}$ symmetry breaking threshold $\lambda_{\mathcal{PT}}$ versus $2 \pi \beta$ for $V_0 / \hbar=3$.}
\end{figure}
   \begin{figure}[b]
\includegraphics[width=8.2cm]{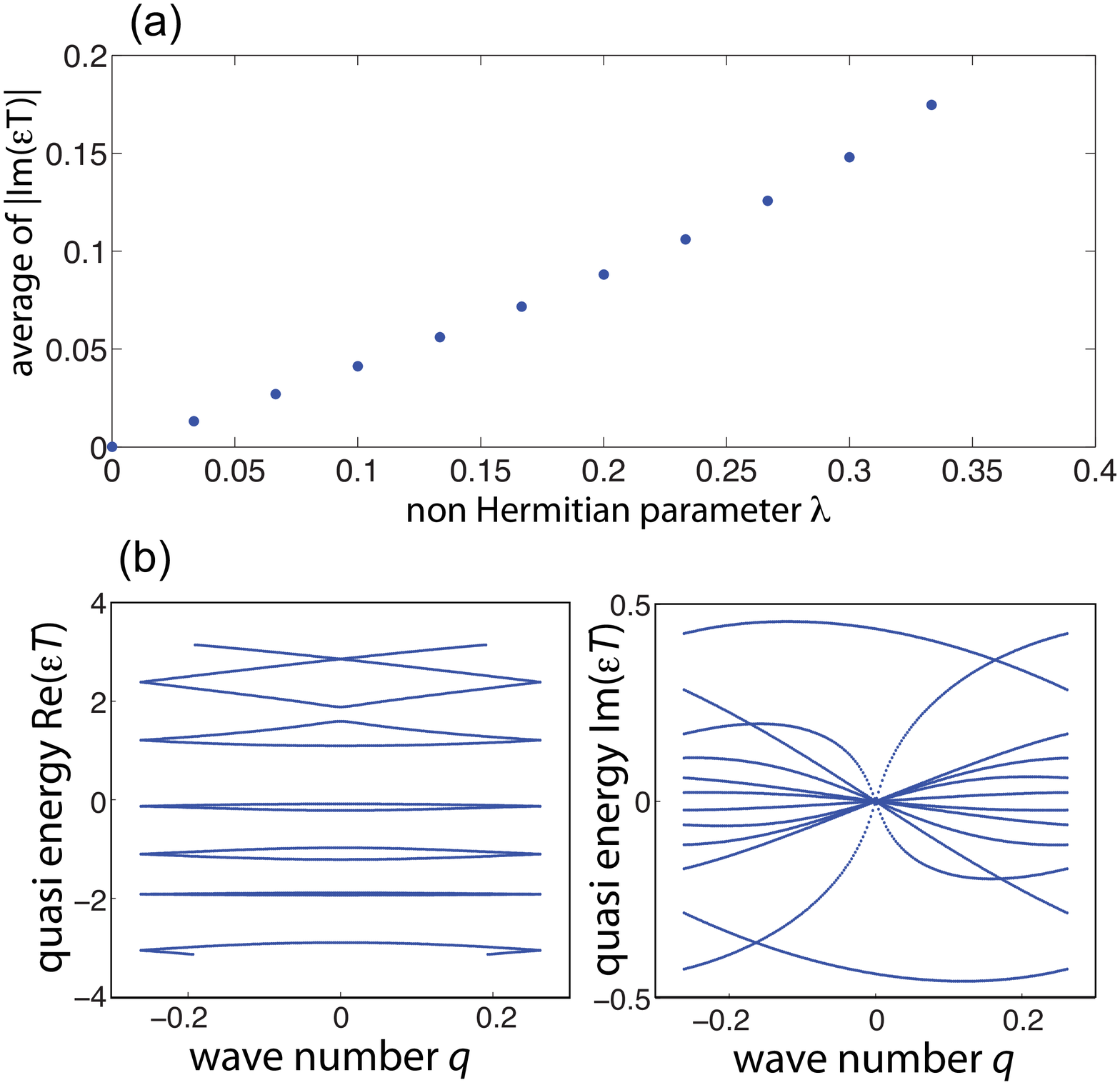}
\caption{Quasi energy spectrum $\epsilon$ in the rational $\beta$ case. (a) Numerically-computed behavior of the average of $|{\rm Im} (\epsilon T)|$ versus $\lambda$ for parameter values $V_0 / \hbar=3$ and $ \beta=1/12$. $\mathcal{PT}$ symmetry breaking is clearly observed at $\lambda_{\mathcal{PT}}=0$. (b) Detailed behavior of the $M=12$ quasi-energy bands (real and imaginary parts of $\epsilon T$) for $V_0 / \hbar=3$ and $\lambda=0.3$.}
\end{figure}
 For $\lambda>0$, owing to the non-Hermitian nature of $\hat{H}$ the quasi energy spectrum is generally complex. Like in the unmodulated case $f(t)=1$, the $\mathcal{PT}$ KR will be said to be in the unbroken $\mathcal{PT}$ phase whenever the entire quasi energy spectrum $\epsilon$ is real, whereas it is said to be in the broken $\mathcal{PT}$ phase if complex conjugate quasi energies appear. As for the Hermitian limit $\lambda \rightarrow 0$, a different behavior is found for rational and irrational values of $\beta$. For a rational value of $\beta=N/M$ the energy spectrum is absolutely continuous and composed by a set of $M$ quasi energy bands (see Appendix A for more details). On the other hand, DL is still observed for a generic irrational $\beta$. Rather generally, by increasing $\lambda$ above $\lambda=0$, a threshold value $\lambda_{\mathcal{PT}}$ is found above which the KR enters into the broken $\mathcal{PT}$ phase. Extended numerical simulations provide strong evidence that, while for  $\beta=N/M$ rational $\lambda_{\mathcal{PT}}=0$, i.e. the KR is always in the broken $\mathcal{PT}$ phase, DL for a generic irrational value $\beta$ assists the $\mathcal{PT}$ phase, i.e. $\lambda_{\mathcal{PT}}>0$ with $\lambda_{\mathcal{PT}}$ increasing toward $\lambda_{\mathcal{PT}}=1$ as $\beta \rightarrow 0$. Typical examples of numerical results, for  irrational and rational values of $\beta$, are shown in Figs.1 and 2, respectively. Figure 1(a) shows the numerically-computed behavior of the average value of $|{\rm Im} (\epsilon T)|$ versus  the non-Hermitian parameter $\lambda$ for the irrational value $2 \pi \beta=0.7$. The two insets in the figure depict the detailed behavior of the real and imaginary parts of the quasi energies $\epsilon T$ of the various Floquet eigenmodes. A value $N_0=1000$ has been typically used in the numerical simulations, corresponding to $2N_0+1=2001$ Floquet modes $|\phi \rangle$. Unphysical Floquet states localized at the edges $l= \pm N_0$, that arise from matrix truncation, have been disregarded in the analysis.
 Figure 1(a) clearly indicates the existence of a $\mathcal{PT}$ symmetry  breaking threshold at $\lambda_{\mathcal{PT}} \simeq 0.27$. Figure 1(b) shows the numerically-computed symmetry breaking threshold $\lambda_{\mathcal{PT}}$ for a few increasing irrational values of $\beta$. Note that, as $\beta$ increases, a lowering of the symmetry breaking threshold is observed, while $\lambda_{\mathcal{PT}} \rightarrow 1^-$ as $\beta \rightarrow 0$. Figure 2 shows, for comparison, a typical result obtained for a rational value of $\beta$, namely $\beta=1/12$. In this case, the average of $|{\rm Im} (\epsilon T)|$ versus $\lambda$ [Fig.2(a)] is an increasing function of $\lambda$ and does not show any threshold, i.e. $\lambda_{\mathcal{PT}}=0$. A typical behavior of the real and imaginary parts of the $M=12$ quasi energy bands $\epsilon T$ at $\lambda=0.3$ are shown in Fig.2(b). Like in the Hermitian KR model, an exception occurs at $\beta=1/2$, corresponding to a quantum anti-resonance: in this case the two quasi energy bands are flat and entirely real, i.e. the KR is always in the unbroken $\mathcal{PT}$ phase (see Appendix A for more details).\\
The numerical results provide strong indication that DL assists the $\mathcal{PT}$ unbroken phase of the quantum KR, while diffusive behavior at quantum resonances shortly brings the $\mathcal{PT}$ KR in the broken $\mathcal{PT}$ phase. The important result that DL helps to preserve the unbroken $\mathcal{PT}$ phase can be qualitatively explained as follows. For $\lambda<1$, the complex sinusoidal potential (3) can be written in the form
\begin{equation}
V(x)=K_0 \cos [ 2 \pi (x-ix_0)/a ]
\end{equation} 
 where $K_0$ and $x_0$ are defined by the relations 
  \begin{equation}
 K_0 \equiv V_0 \sqrt{1-\lambda^2} \;\; , \; \; {\rm tanh} ( 2 \pi x_0/a)  \equiv \lambda. 
 \end{equation}
 Let us indicate by $\hat{H}_0$ the Hermitian KR Hamiltonian
 \begin{equation}
 \hat{H}_0 = -\frac{\hbar^2}{2I} \frac{\partial^2}{\partial x^2}+ K_0 \cos (2 \pi x /a) \sum_n \delta (t-nT)
 \end{equation}
 and let us assume that $\beta$, defined by Eq.(5), is irrational. Then it is obvious that, if $\psi_0(x,t)$ is a solution to the Schr\"{o}dinger equation $i \partial_t \psi_0= \hat{H}_0 \psi_0$, then the complex displaced function $\psi(x,t)=\psi_0(x-ix_0,t)$ is a solution to the Schr\"{o}dinger equation $i \partial_t \psi= \hat{H} \psi$, with $\hat{H}$ defined by Eqs.(1-3). In particular, this implies that, if $ | \theta  \rangle = \sum_l \theta_l \exp(2 \pi i l x /a) $ is a localized Floquet eigenstate of $\hat{H}_0$ with quasi energy $\epsilon$  and localization length $\xi_L$, then $ | \phi \rangle = \sum_l \phi_l \exp( 2 \pi i l x/a)$ is a Floquet eigenstate of $\hat{H}$ with the same quasi energy $\epsilon$ once we formally take
 \begin{equation}
 \phi_l=\theta_{l} \exp( 2 \pi l x_0 /a)
 \end{equation}
 and provided that the localization condition
 \begin{equation}
 |\phi_l|= | \theta_l| \exp( 2 \pi l x_0 /a) \rightarrow 0
 \end{equation}
  as $l \rightarrow \pm \infty$ is met. We note that, for a small non-Hermitian parameter $\lambda$, from Eq.(13) one has $K_0 \sim V_0$ and $ 2 \pi x_0/a \sim \lambda $. Since $|\theta_l| \sim \exp(-|l-l_0|/ \xi_L)$ with $\xi_L \sim V_0^2/ (4 \hbar^2)$ [see Eq.(11)], the localization condition (16) implies $ 2 \pi x_0/a < 1/ \xi_L$, i.e. $\lambda < \sim  4 \hbar^2 /V_0^2$. Therefore, for a sufficiently  small yet non-vanishig value of $\lambda$ the localized Floquet eigenstates of the Hermitian KR Hamiltonian $\hat{H}_0$ can be mapped into localized Floquet eigenstates of the $\mathcal{PT}$ KR Hamiltonian $\hat{H}$ with the same energy spectrum. Hence for sufficiently small $\lambda$ localized states of the Hemitian KR model can be mapped onto localized states of the non-Hermitian KR model with the same quasi energy spectrum, i.e. the perturbation $\lambda$ in the potential does not change the energy spectrum. \par
  
  \begin{figure}[b]
\includegraphics[width=8.2cm]{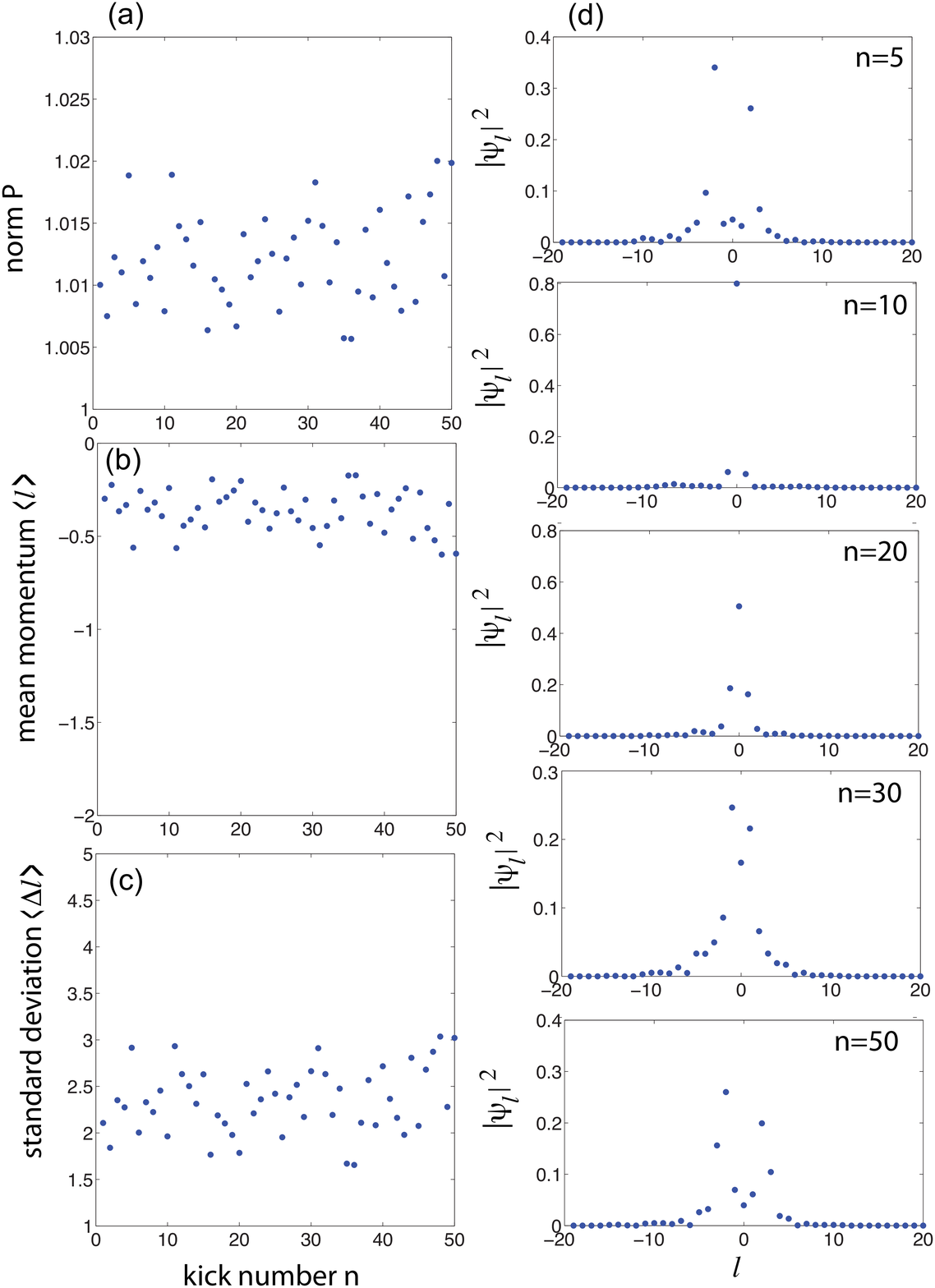}
\caption{Evolution of the momentum distribution $|\psi_l(n)|^2$ versus kick number $n$, for an initial state with zero momentum, in the $\mathcal{PT}$ KR model for $V_0 / \hbar=3$, $\lambda=1/30$ and $ \beta=1/ 4 \pi$, corresponding to dynamic localization and unbroken $\mathcal{PT}$ phase. (a) Behavior of the norm $P(n)=\sum_l |\psi_l(nT)|^2$. (b), (c) Behavior of the mean value $ \langle l (n) \rangle$ and standard deviation $\langle \Delta l (n) \rangle$  of the momentum distribution. (d) Detailed behavior of the momentum distribution $|\psi_l|^2$ at a few kick numbers $n$.}
\end{figure}

\begin{figure}[b]
\includegraphics[width=8.2cm]{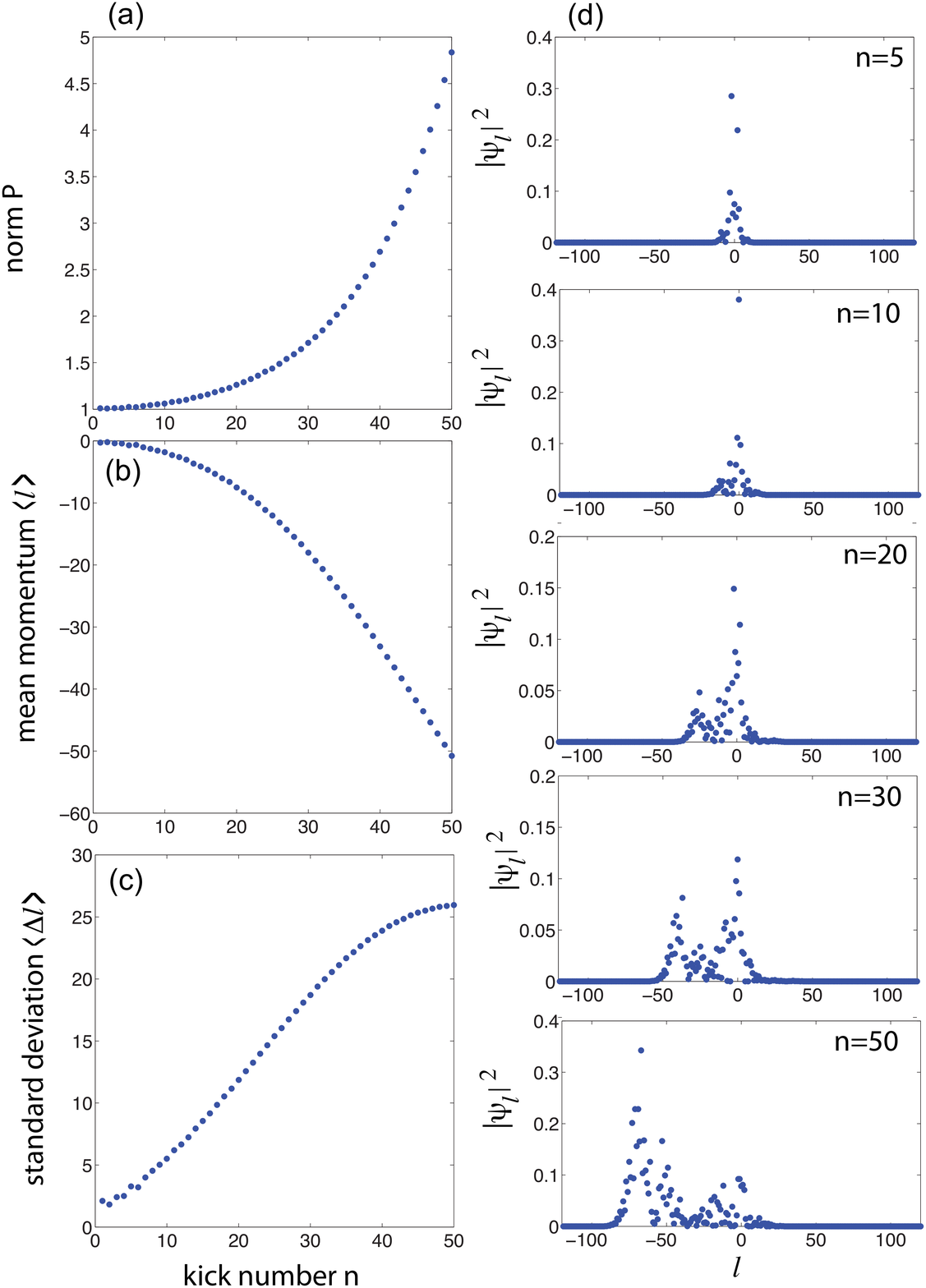}
\caption{Same as Fig.3, but for $ \beta=1/ 12$, corresponding to quantum resonance and broken $\mathcal{PT}$ phase.}
\end{figure}

\section{Quantum resonances and non-Hermitian ratchet acceleration}
The different behavior of the $\mathcal{PT}$ KR for rational (quantum resonances) and irrational (dynamic localization) values of $\beta$ can be seen by considering the temporal evolution of  the particle wave function in momentum space for a zero-momentum initial state. Figures 3 and 4 show typical examples of the dynamics as obtained by numerical integration of Eq.(4) with the initial condition $\psi_l(0)=\delta_{l,0}$ in the dynamic localization ($ \beta=1/4 \pi$) and quantum resonance ($\beta=1/12$) regimes. 
The figures show the evolution of the norm $P(n)= \sum_l |\psi_l(t=nT)|^2$, mean value $ \langle l(n) \rangle = (\sum_l l |\psi_l(nT)|^2)/P$ and standard deviation
$\langle \Delta l(n) \rangle =\sqrt{\sum_l (l-\langle l \rangle)^2 |\psi_l|^2 /P}$
of momentum distribution versus the kick number $n$, as well as the entire distribution $|\psi_l(nT)|^2$ for a few values of $n$. In the simulations the complex sinusoidal potential (3) with $V_0 / \hbar=3$ and $\lambda=1/30$ has been assumed, corresponding to the unbroken (broken) $\mathcal{PT}$ phase for irrational (rational) $\beta$. In the DL regime with unbroken $\mathcal{PT}$ phase (Fig.3), the norm $P(n)$ is bounded, the mean momentum shows small deviations around the initial value $l=0$ and the momentum distribution clearly shows dynamical localization. Conversely, in the quantum resonance regime (Fig.4) the norm $P(n)$ shows a secular (unbounded) growth as a signature of the broken $\mathcal{PT}$ phase and the momentum distribution spreads (diffuses) as the number of kicks $n$ increases. Remarkably, the mean momentum $\langle l(n) \rangle$ shows a secular growth with $n$ [Fig.4(b)], indicating the existence of a ratchet acceleration.  Such an acceleration is a signature of {\it non-Hermitian} transport since it would vanish in the $\lambda=0$ (Hermitian) limit. We can gain physical insights into the appearance of non-Hermitian ratchet acceleration by considering the simplest case of the fundamental quantum resonance $\beta=1$, corresponding to a single quasi energy band, however a similar argument holds {\it mutatis mutandis} by considering other quantum resonances. For $\beta=1$, from Eq.(7) it readily follows that the operator $\exp \left( i \frac{\beta a^2}{ 2 \pi}  \frac{\partial^2}{\partial x^2}  \right)$ on the right hand side of the equation, acting on the space of periodic functions with spatial period $a$, reduces to the identity operator, so that the propagator of the system over the time interval $T$ reads
\begin{equation}
\hat{U}=\exp \left[-\frac{ i V(x)}{ \hbar} \right].
\end{equation}
In momentum space this means that, at the stroboscopic times $t=0,T,2T,..., nT,...$, the amplitude probabilities $\psi_l(t)$ can be obtained as the solutions of the equivalent set of coupled equations (which replace Eq.(4))
\begin{equation}
i \hbar T \frac{d \psi_l}{dt}=\sum_n V_{l-n} \psi_n.
\end{equation}
 For the complex sinusoidal potential [Eqs.(3), (12) and (13)], such equations read explicitly
 \begin{equation}
 i \frac{d \psi_l}{dt}= \frac{K_0}{2 \hbar T} \exp( - 2 \pi x_0/a)  \psi_{l+1}+ \frac{K_0}{2 \hbar T} \exp(  2 \pi x_0/a)  \psi_{l-1}
 \end{equation}
which formally describes the hopping dynamics of a quantum particle on a one-dimensional tight-binding lattice in the presence of an imaginary gauge field \cite{Hatano}. The quasi energy spectrum $\epsilon(q)$ s readily found by setting $\psi_l(t) \sim \exp(iql-i \epsilon t)$ in Eq.(19), yielding 
\begin{eqnarray}
\epsilon(q) & = &  \frac{K_0}{\hbar T} \cos(q+2 \pi i x_0/a) \\
\ & = & \frac{K_0 \cosh ( 2 \pi x_0/a)}{ \hbar T} \cos q-i \frac{K_0 \sinh( 2 \pi x_0/a)}{ \hbar T} \sin q. \nonumber
\end{eqnarray}
where $q$ is the Bloch wave number which varies in the range $(-\pi/a, \pi/a)$. The imaginary gauge field introduces asymmetry in left/right hopping, with preferred transport in one direction \cite{r57,r58,r60}. For the initial condition $\psi_l(0)=\delta_{l,0}$, the solution to Eq.(19) is given by
\begin{equation}
\psi_l(t)= \frac{1}{2 \pi}\int_{-\pi}^{\pi}dq \exp[iql-i \epsilon(q) t]
\end{equation}
which can be written in terms of Bessel function $J_l$ with complex argument. For our purposes, it is sufficient considering the asymptotic behavior of $|\psi_l(t)|^2$ as $t \rightarrow \infty$. In such a limit, the dominant term of the integral on the right hand side of Eq.(21) comes from $q \sim q_0=-\pi/2$, since at $q=q_0$ the imaginary part of the quasi energy $\epsilon(q)$ shows its maximum, i.e. Bloch waves with wave number $q \sim q_0$ show the largest amplification rate.  As shown in the Appendix B, for $t \rightarrow \infty$ one has
\begin{equation}
|\psi_l(t)|^2 \sim \frac{1}{2\pi | \epsilon^{{\rm "}}| t} \exp [ 2 {\rm Im} ( \epsilon(q_0) ) t ] \exp \left[-\frac{(l-v_g t)^2}{| \epsilon^{\rm{"}}| t }  \right] 
\end{equation}
where we have set $v_g \equiv =(d \epsilon / dq)_{q_0}$ (group velocity) and $ \epsilon^{\rm{"}} = (d^2 \epsilon / dq^2)_{q_0}$. Equation (22) clearly shows that in momentum space the distribution $| \psi_l(t)|^2$ is amplified while drifting with a group velocity $v_g$. Therefore, asymptotically one has $\langle l (t)\rangle \sim v_g t$, which explains the appearance of ratchet acceleration according to the results of Fig.4(b). Interestingly, besides a drift the distribution $| \psi_l(t)|^2$ broadens with time like $\Delta l (t) \sim \sqrt{t}$, which differs than the ballistic broadening law $\Delta l (t) \sim t $ found in the Hermitian limit.  

\begin{figure}
\includegraphics[width=8.2cm]{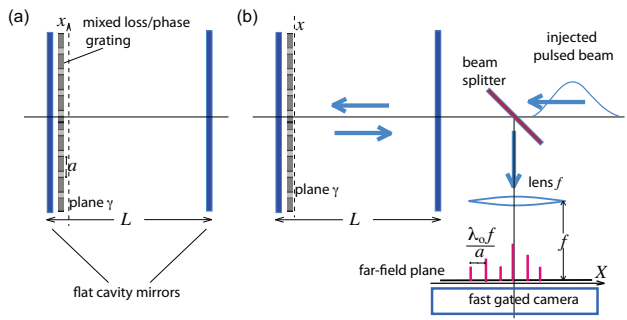}
\caption{(a) Schematic of a Fabry-Perot resonator with flat end mirrors, spaced by $L$, with combined thin index and loss gratings. Beam propagation back and forth between the two end mirrors emulates the $\mathcal{PT}$-symmetric KR Hamiltonian. (b) Schematic setup of excitation of the passive resonator by a pulsed Gaussian beam and recording of far-field intensity images at successive round trips (plane $X$). The far-field intensity images map the $\mathcal{PT}$ KR dynamics in momentum space.}
\end{figure}
\begin{figure}
\includegraphics[width=7.4cm]{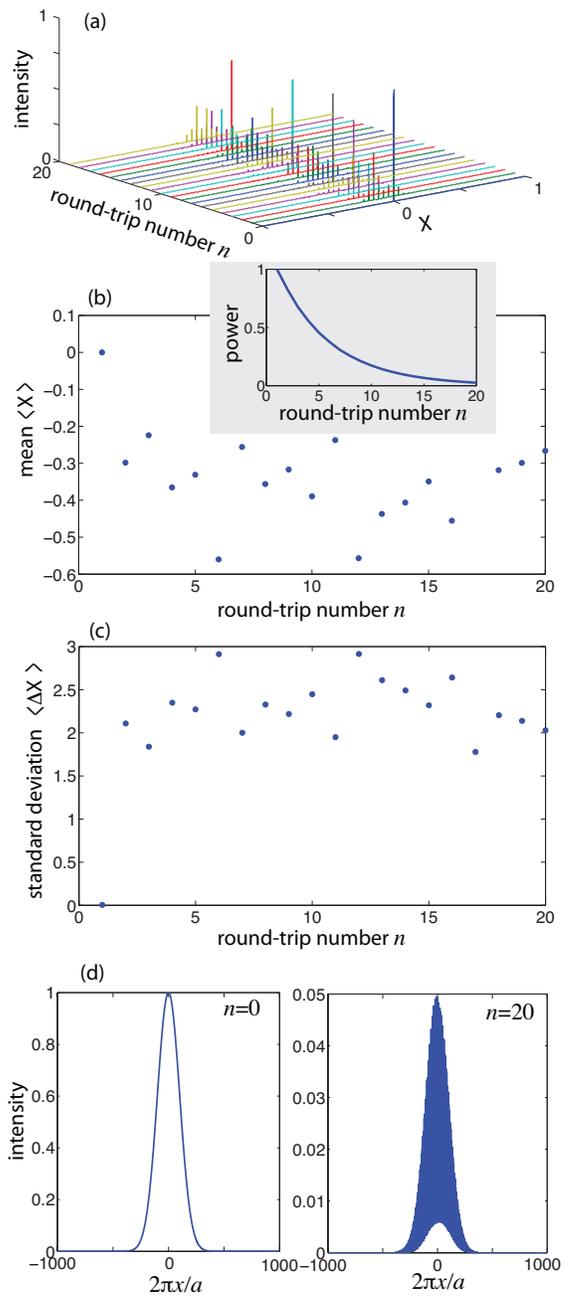}
\caption{Numerical simulations of beam evolution in the resonator of Fig.5(b) for $\beta=1/ (4 \pi)$, after initial excitation with a broad Gaussian beam. Other parameter values are given in the text. (a) Behavior of the intensity distribution (in arbitrary units) at the far-field (focal) plane $X$ at successive round trips (kicks). The peaks on the $X$ axis are spaced by $\lambda_0 f/a$. (b,c) Behavior of the mean value $\langle X \rangle $ and standard deviation $\langle \Delta X \rangle$ of the far-field intensity distribution versus round trip number, in units of $ \lambda_0 f/a$. The inset in (b) shows the evolution of the optical beam power, in arbitrary units, at successive transits in the resonator. (d) Intensity beam distribution at plane $x$ (near-field plane) at $n=0$ (excitation Gaussian beam, left panel) and after $n=20$ round trips (right panel).}
\end{figure}
\begin{figure}
\includegraphics[width=7.4cm]{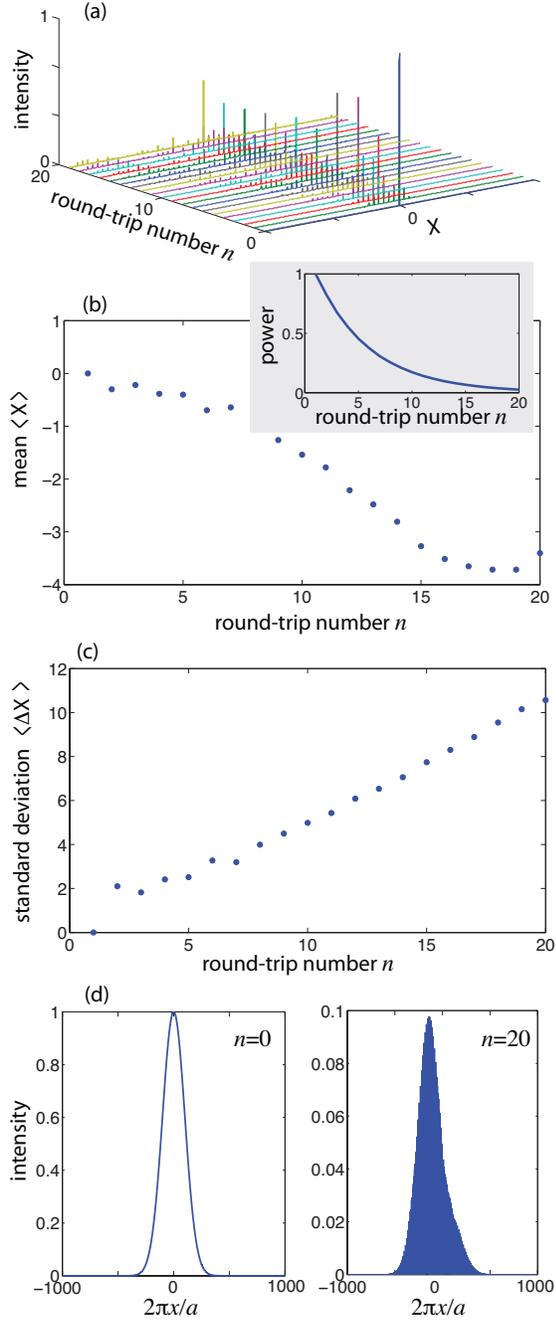}
\caption{Same as Fig.6, but for $\beta=1/ 12$.}
\end{figure}

\section{$\mathcal{PT}$ Kicked Rotor model in a passive optical resonator}
In this section we suggest a simple optical system which can emulate wave dynamics in the $\mathcal{PT}$-symmetric extension of the KR Hamiltonian. In the Hermitian limit, it was shown that paraxial light propagation along a periodic sequence of free-space propagation and phase gratings emulate the KR model \cite{r14,r14bis}, with the observation of DL near a quantum antiresonance. The main limitations of the setup of Refs.\cite{r14,r14bis} is that is requires accurate alignment of the gratings and can be used to observe the dynamical behavior over a relatively low number of kicks.  Here we consider a different setup, namely a passive Fabry-Perot optical resonator with intracavity phase and loss gratings in the setting shown in Fig.5(a). Transverse light propagation back and forth between the two flat cavity mirrors  of the resonator is equivalent to the propagation along a periodic sequence of free diffraction and lumped phase/loss gratings. As compared to the optical KR setup of Refs.\cite{r14,r14bis}, it does not require precise alignment of successive gratings in the sequence and enables to study the beam propagation over a larger number of kicks (round trips in the resonator). The optical resonator is initially excited by a short pulsed optical beam, with pulse duration comparable or shorter than the photon transit time $T_R$ in the cavity, and the successive field decay dynamics in the cavity is investigated in the far field (Fourier plane), as shown in Fig.5(b). Indicating by $\psi_n(x)$ the intracavity field at the reference plane $\gamma$ and at time $t=nT_R$, propagating from the left to the right side, the evolution of the cavity field at successive transits in the cavity is governed by the following map
\begin{equation}
\psi_{n+1}(x)=\exp[-i \theta_1(x)-\theta_2(x)] \exp\left(  i \frac{L}{k_0} \frac{\partial^2}{\partial x^2} \right) \psi_n(x)
\end{equation} 
where $\theta_1(x)$ and $\theta_2(x)$ are the spatial profiles of the phase and loss gratings, respectively, $L$ is the length of the resonator, $k_0= 2 \pi / \lambda_0$, and $\lambda_0$ is the wavelength of the circulating light beam. For sinusoidal and quarter-wave shifted gratings, one can assume
\begin{equation}
\theta_1(x)= A \cos ( 2 \pi x /a) \; , \; \; \theta_2(x)= A \lambda[1-\sin (2 \pi x /a)]
\end{equation}
so that Eq.(23) can be written in the form
\begin{equation}
\psi_{n+1}(x)=\exp[-iV(x)/ \hbar -\gamma] \exp\left(  i \frac{\beta a^2}{ 2 \pi}  \frac{\partial^2}{\partial x^2} \right) \psi_n(x)
\end{equation} 
where we have set
\begin{eqnarray}
\beta & = & \frac{\lambda_0 L}{ a^2} \\
V(x) / \hbar & = & A \left[ \cos( 2 \pi x /a)+i \lambda \sin (2 \pi x / a) \right]\\
\gamma & = A \lambda & 
\end{eqnarray}
A comparison of Eqs.(7) and (25) yields
\begin{equation}
\psi_{n+1}(x)= \exp(-\gamma) \hat{U} \psi_{n}(x)
\end{equation}
where $\hat{U}$ is defined by Eq.(7). This means that beam propagation at successive transits in the resonator, back and forth between the two mirrors, reproduces the dynamics of the $\mathcal{PT}$ KR Hamiltonian after each successive kick, apart from an exponential loss term $\exp(-\gamma)$ that describes a global decay of the field envelope in the passive resonator \cite{note}. Note that the distance $L=L_T$ between the two mirrors that realizes the principal quantum resonance $\beta=1$, namely
\begin{equation}
L_T=\frac{a^2}{\lambda_0}
\end{equation}
corresponds to the well-known Talbot distance \cite{Talb1,Talb2,Talb3}. Since $\beta=L/L_T$, a quantum resonance is thus obtained when the mirror spacing $L$ is set at a fractional Talbot distance.\\
To show the ability to observe the different dynamical regimes for rational and irrational values of $\beta$, including the appearance of non-Hermitian ratchet acceleration, we numerically simulated the decay dynamics of a broad Gaussian beam $\psi_0(x)=\exp(-x^2/w_0^2)$, initially exciting the passive optical cavity of Fig.5, for parameter values $A=3$, $\lambda=1/30$, $w_0/a=100/ \pi$ and for two different resonator lengths, corresponding to $\beta=L/L_T=1/(4 \pi)$ (Fig.6) and $\beta=L/L_T=1/12$ (Fig.7). Recording the field intensity distribution in the focal (far field) plane $X$ of Fig.5(b) at successive round trips enables to reproduce the evolution in momentum space of the $\mathcal{PT}$ KR dynamics. The far-field intensity distribution at plane $X$ is made of a sequence of peaks, spaced one another by $ \lambda_0 f/a$ [see Fig.5(b)], which reproduces the momentum distribution $|\psi_l|^2$ at successive kicks. In an experiment, the instantaneous transverse light distribution in the far field at successive transits in the cavity can be detected using a fast gated camera, as demonstrated e.g. in Refs.\cite{rE1,rE2}.
In physical units, assuming for instance a probing wavelength $\lambda_0=780$ nm (Ti:shappire laser) and a grating period $a=300 \; \mu$m, the Talbot distance is $L_T \simeq 11.54$ cm, the spot size of the broad excitation Gaussian beam is $w_0 \simeq 9.55$ mm, and the mirror spacing in the simulations of Figs.6 and 7 is $ L \simeq 9.182$ mm and $L \simeq 9.615$ mm, respectively.  For a focusing lens $f=5$ cm, the intensity peaks in the far-field plane $X$ are spaced by $\lambda_0 f/a \simeq 130 \; \mu$m.\\
The simulation of Fig.6 corresponds to the irrational case $\beta=1/ ( 4 \pi)$. Figure 6(a) shows the evolution of the far-field intensity distribution (in arbitrary units) at successive round trips $n$, whereas the evolution of the center of mass  $\langle X \rangle$ and standard deviation $\langle \Delta X \rangle$ of the distribution, in units of $ \lambda_0 f/a$, is shown in Figs.6(b) and (c). Clearly DL is observed, whereas ratchet acceleration is not present. The inset in Fig.6(b) shows the evolution of the optical beam power (in arbitrary units) at successive transits, showing an exponential decay that arises from the losses in the absorptive grating. Finally, Fig. 6(d) shows the intensity field distribution at plane $x$ (near-field plane) of the excitation Gaussian beam (left panel) and of the field after $n=20$ round trips (right panel).   Figure 7 shows the numerical results for the rational case $\beta=1/12$, corresponding to a quantum resonance. In this case there is not DL [Fig.7(c)], while acceleration ratchet is clearly observed [Fig.7(b)]. 

\section{Conclusion}
In this article a $\mathcal{PT}$-symmetric extension of the celebrated quantum KR model has been theoretically investigated, and the role of non-Hermitian dynamics on  wave transport has been highlighted. One of the main result of the analysis if that the phenomenon of dynamical localization, i.e. suppression of quantum diffusion in momentum space, assists the unbroken $\mathcal{PT}$ phase. On the other hand, in the delocalization (quantum resonance) regime the $\mathcal{PT}$ symmetry is always in the broken phase. Remarkably, in the latter case a ratchet acceleration arises as a signature of unidirectional non-Hermitian transport, and quantum diffusion is slower as compared to the ordinary (Hermitian) KR model (spreading law $ \sim \sqrt{t}$ versus $\sim t$ for the main quantum resonance). In the last part of the article we have also suggested 
an optical implementation of the periodically-kicked $\mathcal{PT}$-symmetric Hamiltonian, based on transverse light dynamics in a passive optical resonator with intracavity phase and loss (absorptive) gratings. In particular, by tuning the cavity length at a fractional Talbot distance, non-Hermitian ratchet acceleration should be clearly observable by monitoring the light beam evolution in a far-field plane. \\
Our results disclose important physical effects of wave transport in non-Hermitian and $\mathcal{PT}$-symmetric models that are envisaged to stimulate further theoretical and experimental investigations. In particular, our results highlight the interplay between  dynamical localization and $\mathcal{PT}$ symmetry, and suggest a new way to realize ratchet acceleration based on non-Hermitian transport.

\appendix
\section{Quantum resonances and quasi-energy bands}
Like for the Hermitian KR Hamiltonian, a quantum resonance for the $\mathcal{PT}$-symmetric KR model arises when $\beta$ is a rational number, i.e. $\beta=N/M$ with $N \leq M$ and $N,M$ coprime integers. In this case the quasi energy spectrum of $\hat{H}$ is absolutely continuous with $M$ quasi energy bands and delocalized Floquet eigenstates. In fact, when $\beta=N/M$ the Floquet eigenstates $\psi_l$, satisfying the linear system (8), can be searched in the form
\begin{equation}
\phi_l=c_l \exp(iql)
\end{equation}
where $q$ is an integer number that varies in the range $ -\pi/M \leq q < \pi/M$ and 
\begin{equation}
c_{l+M}=c_l.
\end{equation}
 Substitution of Eq.(A.1) into Eq.(8) and using the property that $\mathcal{U}_{n+M,l+M}=\mathcal{U}_{n,l}$ [see Eq.(9)], it readily follows that Eq.(8) is satisfied provided that the $M$ amplitudes $c_0$, $c_1$, ..., $c_{M-1}$ satisfy the linear homogeneous set of equations
\begin{equation}
\exp(-i \epsilon T) c_l= \sum_{n=0}^{M-1} S_{l,n}c_n
\end{equation}
where the $M \times M$ matrix coefficients $S_{l,n}=S_{l,n}(q)$ are defined by
\begin{equation}
S_{l,n}(q)= \sum_{\alpha=-\infty}^{\infty} W_{l-\alpha M -n} \exp[iq(\alpha M+n-l)-2 \pi i \beta n^2]
\end{equation}
Hence the quasi energies $\epsilon$, corresponding to delocalized Floquest eigenstates Eqs.(A1) and (A2), are found from the eigenvalues $\exp(-i \epsilon T)$ of the $M \times M$ matrix $S_{l,n}(q)$ according to Eq.(A3). Since $q$ varies in the range $-\pi/M \leq q < \pi/M$, the quasi energies $\epsilon=\epsilon(q)$ are composed by $M$ bands.\par
Since the quasi energy spectrum is continuous and the corresponding Floquet eigenstates are delocalized, delocalization and ratted acceleration in momentum space is generally observed for a particle with initial zero momentum. The special case corresponding to the main quantum resonance $\beta=1$  is discussed in details in Appendix B. Moreover, in the non-Hermitian case the imaginary parts of the quasi energies are generally nonvanishing, indicating that the $\mathcal{PT}$ symmetry is always broken in the rational $\beta$ case. An exception occurs at $\beta=1/2$, so-called quantum antiresonance, where the $M=2$ quasi energy bands are flat, localization is found and the $\mathcal{PT}$ symmetry is unbroken. One can readily prove such a property by considering the form of the propagator $\hat{U}$, given  by Eq.(7) in the text, in the special case $\beta=1/2$.  Since $\hat{U}$ acts on the space of functions which are periodic in space with spatial period $a$, the operator $\exp \left( i \frac{\beta a^2}{ 2 \pi}  \frac{\partial^2}{\partial x^2}  \right)
$ for $\beta=1/2$ is equivalent to the translation operator $\hat{T}_{a/2}$, with $\hat{T}_{a/2} \psi(x)=\psi(x+a/2)$. Hence
\begin{eqnarray}
\hat{U}^2 & = & \exp \left[-\frac{ i V(x)}{ \hbar} \right]  \hat{T}_{a/2} \exp \left[-\frac{ i V(x)}{ \hbar} \right]  \hat{T}_{a/2} \nonumber \\
& = & \exp \left[-i\frac{  V(x)+V(x+a/2)}{ \hbar} \right] \hat{T}_a =1
\end{eqnarray}
since $\hat{T}_a=1$ and $V(x)+V(x+a/2)=0$. This means that, after two kicks, the system returns to its initial state. Since $\hat{U}^2 | \phi \rangle= \exp( -2 i \epsilon T) | \phi \rangle$, two values of quasi energies, $\epsilon T=0, \pi$, are thus allowed, corresponding to two flat bands with real spectrum.

\section{Asymptotic behavior of momentum distribution at the main quantum resonance $\beta=1$}
In this Appendix we calculate the asymptotic behavior of $\psi_l(t)$ for the main quantum resonance $\beta=1$, Eq.(21) given in the text, for long times $t$. To this aim, let us note that as $ t \rightarrow \infty$ the main contribution to the integral on the right hand side of Eq.(21) comes from the wave numbers $q$ close to $q_0= -\pi/2$, where the imaginary part of the dispersion curve $\epsilon(q)$ [Eq.(20)] has its maximum. We can therefore expand $\epsilon(q)$ around $q=q_0$ up to second order by letting
\begin{equation}
\epsilon(q) \simeq \epsilon(q_0)+ v_g (q-q_0)+\frac{1}{2} \epsilon^{\rm{"}} (q-q_0)^2
\end{equation}
where we have set $v_g \equiv (d \epsilon / d q)_{q_0}$ and $\epsilon^{\rm{"}} \equiv (d^2 \epsilon / d q^2)_{q_0}$. Note that $v_g$ is real, while $\epsilon^{\rm{"}}$ is imaginary with ${\rm Im}( \epsilon^{\rm{"}})<0$. Substitution of Eq.(B1) into Eq.(21) yields
\begin{eqnarray}
\psi_l(t) & \sim & \frac{1}{2 \pi} \exp \left[ i q_0 l-i \epsilon(q_0) t \right] \times \nonumber \\
& \times &  \int d \delta \exp \left[ i \delta (l-v_gt) -\frac{1}{2} | \epsilon^{\rm{"}}| \delta^2 t \right]
\end{eqnarray} 
where $\delta=q-q_0$. For large $t$, the Gaussian function under the sign of the integral on the right hand side of Eq.(B2) gets narrower around $\delta=0$, so that the integral can be extended from $-\infty$ to $\infty$ and calculated in a  closed form (generalized Gaussian integral). This yields
\begin{equation}
\psi_l(t) \sim \frac{1}{\sqrt{2 \pi | \epsilon^{\rm{"}}| t }}   \exp \left[ i q_0 l-i \epsilon(q_0) t \right]  \exp \left[ -\frac{(l-v_gt)^2}{2 |\epsilon^{\rm{"}}|t}\right]
\end{equation}
After taking $| \psi_l(t)|^2$, one finally obtains Eq.(22) given in the text.


\begin{thebibliography}{0}
\bibitem{note}
For a passive cavity with non-vanishing mirror transmittances, $\gamma$ should also include output coupling losses of the resonator.
\end{thebibliography}
\end{document}